\documentclass{tlp}
\pdfoutput=1

\usepackage{xspace}
\usepackage{latexsym}

\usepackage{ifthen}
\newboolean{commentson} %
\setboolean{commentson}{true}
 \setboolean{commentson}{false} %

\newboolean{commentsaon}         %
\setboolean{commentsaon}{true}
 \setboolean{commentsaon}{false}  %

\newcommand{\comment}[1]
{\ifthenelse{\boolean{commentson}}
   {{\par\noindent\mbox{}{\small\blue[ *** #1 ]\par}\noindent\par}}{}}

\newcommand{\commenta}[1]
{\ifthenelse{\boolean{commentsaon}}
   {{\par\noindent\mbox{}{\small\color[rgb]{0, .5, 0}[ *** #1 ]\par}\noindent\par}}{}}

\usepackage{color}
\newcommand\blue     {\color{blue}}

\usepackage{hyperref}
\usepackage{breakurl}

\newtheorem{lemma}
{Lemma}

\newtheorem{definition}
{Definition}

\newtheorem{theorem}
{Theorem}

\newtheorem{corollary}
{Corollary}

\newcommand*{\seq}[2][n]  {{#2_{1}, \allowbreak \ldots, \allowbreak #2_{#1}}}

\newcommand*{\pre}[1]  {{{}^\bullet\!#1}}

\newcommand*{\magic}{{\it magic}}

\newcommand{\notmodels}{\mathrel{\,\not\!\models}}

\newcommand{\LL}{{\ensuremath{\cal L}}\xspace}
\newcommand{\PP}{{\ensuremath{\pre{\cal P}}}\xspace}
\newcommand{\TT}{{\ensuremath{\cal T}}\xspace}
\newcommand{\DD}{{\ensuremath{\cal D}}\xspace}

\newcommand{\ol}{\overline}

\begin{document}

  \submitted{21 December 2009}
  \revised{7 August 2010}
  \accepted{ ? }

\title[Correctness proof, magic transformation]
      {A Simple Correctness Proof \\ for Magic Transformation
      }

\author[W. Drabent]
       {W{\l}odzimierz Drabent\\
         Institute of Computer Science,
         Polish Academy of Sciences,\\
         ul.\ Ordona 21,
         Pl -- 01-237 Warszawa, Poland \\ and \\
         Link\"opings universitet, Department of Computer
         and Information Science\\
         S -- 581\,83   Link\"oping, Sweden      \\
         \email{drabent\,{\it at}\/\,ipipan\,{\it dot}\/\,waw\,{\it dot}\/\,pl}
}

\maketitle

\begin{abstract}
The paper presents a simple and concise proof 
of correctness of the magic transformation.
We believe it may provide
a useful example of formal reasoning about logic programs.

The correctness property concerns the declarative semantics.
The proof, however, refers to the operational semantics (LD-resolution)
of the source programs.
Its conciseness is due to applying a suitable proof method.

\end{abstract}

\section{Introduction}

Magic transformation
(see \cite[Chapter 15.3]{nilsson.maluszynski.book} for references)
is a technique to facilitate efficient bottom-up evaluation of logic programs.
Given a program and an initial goal, the transformation produces
a so-called magic program; 
the answers of both programs for the initial goal should be the same.
Looking for a correctness proof of magic transformation
I found that
such a proof was rather easy to construct.
Moreover the result turned out to be surprisingly concise.
In this note I present the proof with all the details.
I believe it provides a useful example of formal reasoning about logic programs.

Mascellani and Pedreschi \citeNN{DBLP:conf/birthday/MascellaniP02} stated
that 
``all known proofs of correctness of the magic-sets transformation(s)
are rather complicated''
(see \cite{DBLP:journals/jlp/Ramakrishnan91} for an example),
and presented a simpler proof, which concerns the
declarative semantics of the original and transformed programs.
Our proof is maybe even simpler;
moreover it formalizes the relation between the
declarative semantics of the transformed program and the operational
semantics of the original one.
The simplification is due to applying a suitable proof method for
program correctness, instead of constructing a proof from scratch.

\section{Preliminaries}
For standard notions and notation see \cite{Apt-Prolog}.
We consider definite clause programs (not restricted to Datalog).
By a {\em query} we mean a conjunction of atoms.
Given a program $P$, by an {\em answer} (or  {\em correct answer}) we mean
any query $Q$
which is a logical consequence of the program ($P\models Q$).
If an answer is an instance of some initial query $Q_0$ then we say
that it is an answer for $P$ and $Q_0$.
By a  {\em computed answer} for a program $P$ and initial query $Q_0$,
we mean an instance $Q_0\theta$ of $Q_0$,
produced by a successful SLD-derivation for $P$ and $Q_0$.%
\footnote{%
    In  \cite{Apt-Prolog} answers are also called correct instances of
    initial queries, and
    computed answers are called computed instances.
}
A fundamental theorem relates 
answers and computed answers:
\begin{theorem}[Soundness and completeness of SLD-resolution]
\label{th.sound-compl}
\rm
For any program $P$, any query $Q$,
and any selection rule:

If $Q$ is a computed answer for $P$ then $P\models Q$.

If $P\models Q\theta$ then there exists a computed answer $Q\sigma$
for $P$ and $Q$, such that $Q\theta$ is an instance of  $Q\sigma$.%
\footnote{%
    For a proof see e.g.\ \cite[Th.\,4.4, 4,13]{Apt-Prolog}.
}
\end{theorem}

A {\em proof tree}
(called sometimes {implication tree} or {derivation tree})
for a program $P$
and an atomic query $A$
is a finite tree 
whose nodes are atoms,
the root is $A$,
and in which if $\seq B$ ($n\geq0$) are the children of a node $H$ 
then $H\gets\seq B$ is an instance of a clause of~$P$.
Proof trees provide 
a useful characterization of logic program answers:

\begin{theorem} \rm
\label{th.characterization}
For any program $P$ and query $Q$,
$P\models Q$ iff for each atom $A$ of $Q$ there exists a proof tree
for $P$ and $A$.
\end{theorem}
The theorem follows immediately from \cite[Th.\,4.24(v)]{Apt-Prolog}.
The latter is attributed to \cite{Clark79} in
\cite[Proposition 2.6]{DBLP:journals/tcs/Deransart93}.

\medskip
We focus on LD-resolution (SLD-resolution with the Prolog
selection rule) and 
will study the sets of procedure calls and procedure successes in
LD-derivations.
  The procedure calls are the atoms selected in the derivation.
  A definition of procedure successes is given in \ref{appendix}.
  For the proof of the main theorem of this paper it is sufficient to know 
  that any computed answer for an initial atomic query is a procedure success.

Consider a pair ${\langle\it pre,post\rangle}$ of sets of atoms, each
closed under substitution.  We can treat such a pair as a specification
of procedure calls and successes of a program
(a {\em call-success specification}).
\begin{definition}
\rm
We say that a program $P$ with
a query $Q$ is {\em correct} w.r.t.\ 
a call-success specification ${\langle\it pre,post\rangle}$
iff in any LD-derivation for $P$ and $Q$ all the procedure
calls are in ${\it pre}$ and all the successes in  ${\it post}$.
\end{definition}
Notice that such correctness is not a declarative property,
 as it depends on a particular operational semantics.
We will use the following sufficient criterion for  correctness
\cite{DBLP:journals/tplp/DrabentM05}.
(See
 Concluding Remarks for discussion and references, and \ref{appendix}
for a proof.)
\begin{theorem} \rm 
\label{th.correctness}
Assume that for a call-success specification ${\langle\it pre,post\rangle}$,
a program $P$, and an atomic query $Q\in{\it pre}$ the following holds:

For each (possibly non-ground) instance 
$H\gets B_1,\ldots,B_n$ $(n\geq0)$
of each clause of $P$
\begin{equation}\label{VCoperational}
\mbox{}\hspace{-1em}
\begin{array}{l}
\mbox{
if $H\in \it pre, \  B_1,\ldots,B_n \in post$ then  $H\in\it post$,
}
\\
\mbox{
 if $H\in \it pre, \  B_1,\ldots,B_{i-1} \in post$ then  $B_{i}\in\it pre$
 (for $i=1,\ldots,n$)}.
\end{array}
\hspace{-.5em}
\end{equation}
Then $P$ with $Q$ is correct w.r.t.\ 
 ${\langle\it pre,post\rangle}$.
\pagebreak[3]

For a non-atomic initial query the requirement $Q\in{\it pre}$ 
has to be generalized to: for each instance $\seq B$ $(n>0)$ of the query,
 if $  B_1,\ldots,B_{i-1} \in post$ then  $B_{i}\in{\it pre}$
(for $i=1,\ldots,n$).
\end{theorem}

It remains to define the magic transformation.
    It adds new predicate symbols to the alphabet~$\LL$ of 
     programs and queries;
    for each predicate symbol $p$, the unique new symbol $\pre p$ is added.
In a simple version, for instance that of 
\cite{nilsson.maluszynski.book}, the arity of  $\pre p$ is that of $p$.
In the general case, some $k_p$ argument positions of $p$ are
selected, and the arity of $\pre p$ is $k_p$.
(We do not discuss 
the choice of $k_p$ and of the selected positions, as it is irrelevant for
the correctness of magic transformation.)
    Let \PP denote the set of new predicate symbols.
If $A=p(\seq t)$ is an atom over \LL then  $\pre A$ denotes
 $\pre p(t_{i_1},\ldots,t_{i_{k_p}})$,
where $i_1,\ldots,i_{k_p}$ are the selected positions of $p$.
Such an $\pre A$ is called {\em magic template}.
    In what follows $A,B,H$, possibly with subscripts, denote atoms over \LL.
    (Hence  $\pre A, \pre B, \pre H$ stand for atoms with the new
    predicate symbols.)

\begin{definition}[Magic transformation]
\label{def.magic}
\rm
Let $P$ be a program and $Q$ an atomic query.
The magic program $\magic(P,Q)$ for $P$ and $Q$ is the program containing
\begin{enumerate}
\item
\label{magic.modified.clauses} 
a clause $H\gets \pre H, \seq B$ for each clause 
 $H\gets  \seq B$ in $P$,
\item
\label{magic.new.clauses}
a clause $\pre B_i\gets \pre H,  B_1,\ldots,B_{i-1}$ for each clause 
 $H\gets  \seq B$ in $P$ and each $i=1,\ldots,n$,
\item
\label{magic.seed}
the clause $\pre Q\gets$.
\end{enumerate}
\end{definition}
\section{The proof}
\label{sec.proof}

Now we are ready to prove correctness of the magic transformation.
The required property is that both programs have the same answers for $Q$.
Our proof consists of two lemmas (inclusion in two directions).
Moreover, the second lemma formalizes the main intuition
behind the transformation:
program $\magic(P,Q)$ describes the sets of procedure calls and successes in 
computations of program $P$ and query $Q$,
under Prolog selection rule.
In the lemmas, $P$ is a program and $Q$ is an atomic query,
both over \LL.

\begin{lemma}\rm
\label{lemma1}
For any query $R$ over \LL, 
if $\magic(P,Q)\models R$ then $P\models R$.

\end{lemma}

\noindent PROOF:
Consider a proof tree \TT for  $\magic(P,Q)$  and $A$, 
where $A$ is an atom from $R$.
Removing from \TT each atom $\pre B$ results in (a set of trees
containing) a proof tree for $P$ and $A$. 
Thus by Th.\,\ref{th.characterization},
if $\magic(P,Q)\models R$ then $P\models R$.
\hfill $\Box$

\begin{lemma}\rm
\label{lemma2}
$P$ with $Q$ is correct w.r.t.\ a call-success specification 
 ${\langle\it pre,post\rangle}$ given by
\[
\begin{array}{l}
{\it pre} = \{\,  A \mid \magic(P,Q) \models \pre A \,\},
\\
{\it post} = \{\, A \mid \magic(P,Q) \models A \,\}.
\end{array}
\]
In particular, each computed answer $Q\theta$ for $P$ and $Q$ is 
in $post$.
\end{lemma}

\noindent PROOF (outline):
Notice that the magic program is an encoding of the correctness
conditions from Th.\,\ref{th.correctness}. \hfill $\Box$

\pagebreak[3]
\noindent PROOF (detailed):
The magic program contains $\pre Q\gets$, hence $Q\in {\it pre}$.
Consider an instance $(H\gets\seq B)\theta$ of a clause of $P$.
Assume that $H\theta\in {\it pre}$ and 
$B_1\theta,\ldots,B_{i-1}\theta \in post$ ($0<i\leq n+1$).
Then $\magic(P,Q)\models \pre H\theta, B_1\theta,\ldots,B_{i-1}\theta$.
If $i=n+1$ then  $\magic(P,Q)\models H\theta$ (by the clause from case
\ref{magic.modified.clauses} of Def.\,\ref{def.magic}).
If $i\leq n$ then  $\magic(P,Q)\models\pre B_i\theta$ (by the clause from case
\ref{magic.new.clauses} of Def.\,\ref{def.magic}).
Thus the sufficient condition for correctness (from Th.\,\ref{th.correctness})
 is satisfied.
 \hfill $\Box$
\begin{corollary}\rm
\label{corollary.lemma}
If $P\models Q\sigma$ then  $\magic(P,Q)\models Q\sigma$. 
\end{corollary}
\noindent PROOF:
By completeness of LD-resolution, $Q\sigma$ is an instance of a computed
answer $Q\theta$ for $P$ and $Q$.
By Lemma \ref{lemma2}, $Q\theta\in post$.  Hence  $Q\sigma\in post$.  
 \hfill $\Box$

\medskip

From Lemma \ref{lemma1} and the corollary it immediately follows: 

\begin{theorem}[Correctness of the transformation]
\label{th.magic}
\rm
Let $P$ be a program, $Q$ an atomic query,  and $\theta$ a substitution.
Then
\[
\begin{array}{c}
P\models Q\theta \quad  \mbox{ iff } \quad \magic(P,Q) \models Q\theta.  
\end{array}
\]
\end{theorem}

In other words, programs
$P$ and $\magic(P,Q)$ have the same sets of answers for $Q$.
Hence by Th.\,\ref{th.sound-compl},
 any computed answer for $P$,  $Q$ is an instance of a
computed answer for $\magic(P,Q)$,  $Q$\/; and
any computed answer for $\magic(P,Q)$, $Q$ is an instance of a computed answer
 for $P$,  $Q$.
The correctness is sometimes expressed in a less general way, as in the
corollary below
(which follows immediately from Th.\,\ref{th.magic}).

\begin{corollary}\rm
\label{corollary.theorem}
$   {\cal M}_P \cap [Q] =
    {\cal M}_{\magic(P,Q)} \cap [Q] ,
$
where ${\cal M}_P$ denotes the least Herbrand model of $P$, and
 $[Q]$  the set of ground instances of $Q$.
\end{corollary}

\paragraph{\bfseries Variants of magic transformation.}

The reader is encouraged to check that
the proof is also valid for a class of magic transformations, 
characterized as follows:
1.~in a clause $H\gets \pre H,\ldots$ from case
\ref{magic.modified.clauses} of Def.\,\ref{def.magic},
the body atom $\pre H$ may be removed;
2.\mbox{ }some body atom(s) from a clause $\pre B_i\gets\ldots$
(Def.\,\ref{def.magic}, case \ref{magic.new.clauses})
may be removed \cite{nilsson.maluszynski.book}.

In some approaches (e.g.\ \cite{DBLP:journals/jlp/BeeriR91}),
an atom $\pre B_i$ may be added to the body of a magic program clause,
when the body contains $B_i$.  Such program is logically equivalent to
$\magic(P,Q)$, thus our correctness theorem holds also for this case.%
\footnote{%
To show the equivalence, let $P'$ be the program $\magic(P,Q)$
  modified as described.
Any clause of $P'$ can be seen as 
$C' = A\gets\pre H,\seq[i-1]B,F$, where
$C = A\gets\pre H,\seq[i-1]B$ is a clause of  $\magic(P,Q)$, and $F$ is a
possibly empty conjunction of some literals of the form  $\pre B_j$ 
($j<i$). 
Formula $C\to C'$ is a tautology,
hence $\magic(P,Q)\models P'$.

To show $P' \models \magic(P,Q) $,
we prove by induction on $i$ that $P'\models C$,
for each clause  $C\in\magic(P,Q)$ as above.
For $i=1$,  $C=C'\in P'$, as $F$ is empty.
For the inductive step,  assume
without loss of generality that $F$ is a single atom $\pre
B_j$. 
There is a clause $C_{B_j} = \pre B_j\gets\pre H,\seq[j-1]B$ in $\magic(P,Q)$,
where $j<i$. 
By the inductive assumption,
$P'\models C_{B_j}$.
Also,  $P'\models C'$.
Formula $(C_{B_j}\land C')\to C $ is a tautology (e.g.\ apply the resolution
principle w.r.t.\ $F$ to  $C_{B_j}$ and $C'$). 
Thus $P'\models C$.

} %

An important class of magic transformations employs adornments
(see e.g.\ \cite{DBLP:journals/jlp/Ramakrishnan91,DBLP:journals/jlp/BeeriR91}).
The original program $P$ is transformed into an {\em adorned program}
$P^{ad}$, by renaming predicate symbols into fresh ones.  
(We omit the details of the transformation.)
A symbol $p$ may be renamed into more than one symbols; thus several renamings
of a clause $C\in P$ may appear in $P^{ad}$.
Similarly, the query $Q$ is transformed into $\ol Q$
(by applying a selected renaming of its predicate symbol).
The two programs are equivalent in the sense that 
$P\models Q\theta$ iff  $P^{ad}\models\ol Q\theta$.
The new magic program is obtained by applying
the magic transformation from Df.\,\ref{def.magic} to the adorned program:
$
\magic'(P,Q) = \magic(P^{ad},\ol Q),
$
From Th.\,\ref{th.magic} we obtain%
\footnote{%
The proof is: 
$P\models Q\theta$ iff  $P^{ad}\models\ol Q\theta$
iff (by Th.\,\ref{th.magic}) $\magic(P^{ad},\ol Q)$.
}
correctness of this magic transformation:
$P\models Q\theta$ iff $\magic'(P,Q) \models\ol Q\theta$.

\section{Concluding remarks}

We first outline some other correctness proofs of magic transformation.
Then we discuss the method of Th.\,\ref{th.correctness} used in our proof.

Mascellani and Pedreschi  \citeNN{DBLP:conf/birthday/MascellaniP02} prove
 the equivalence
${\cal M}_P \cap [Q] =   {\cal M}_{\magic(P,Q)} \cap [Q]$
of Corollary \ref{corollary.theorem}.
The proof employs Herbrand interpretations.  
In particular it studies the intersection 
of the least Herbrand models (of $\magic(P,Q)$ and of $P$) 
with a Herbrand interpretation $I$, which is related to the set ${\it pre}$ of 
Lemma \ref{lemma2}.

The main part of the proof of
\cite[Th.\,5.1]{DBLP:journals/jlp/Ramakrishnan91},
corresponding to proving Corollary \ref{corollary.lemma},
is based on constructing a proof tree for $\magic(P,Q)$ and $Q$, whenever a
proof tree for $P$ and $Q$ exists.  The proof is by induction on the tree for
$P$.  
The inductive step considers an instance
$Q\gets\pre Q, \seq B  $
of a clause of  $\magic(P,Q)$.
By the inductive assumption, there exist trees for 
$\magic(P,B_i)$ and $B_i$.
To construct trees for $\magic(P,Q)$ and each $B_i$, one needs to show that 
$\magic(P,Q)\models\pre B_i$.  This is done by induction on $i$.
The correctness proof of
\cite{DBLP:journals/jlp/BeeriR91}
is similar.

An important intuition about the magic transformation,
and a motivation for introducing it,
seems to be the correspondence between the magic program and the calls and
successes of the original one.
This correspondence is neglected in the aforementioned proofs.  
In contrast, we formalize it as Lemma~\ref{lemma2}, and it is a core of our
proof.

Nilsson \citeNN{DBLP:journals/tcs/Nilsson95}
presented a concise proof of a property related to Lemma~\ref{lemma2} and Th.\,\ref{th.magic}.
He showed correspondence between the declarative semantics%
\footnote{%
    More precisely, the s-semantics \cite{s-semantics94}.
}
of ${\magic(P,Q)}$
and the collecting top down abstract interpretation of $P$ with $Q$.
The latter provides supersets of the set of calls and the set of successes in
LD-derivations.
So the main idea is similar to that of our proof, however  
 the notion of
abstract interpretation is additionally employed.

The main reason for conciseness of the proof of Th.\,\ref{th.magic}
was employing the correctness proof method of
Th.\,\ref{th.correctness} \cite[Section 3.2]{DBLP:journals/tplp/DrabentM05}.
The method deals with properties of LD-derivations.
Such a property may be non-declarative
(i.e.\ inexpressible by means of the declarative semantics).
The sufficient condition from Th.\,\ref{th.correctness}
was initially proposed
by Bossi and Cocco \citeNN{DBLP:conf/tapsoft/BossiC89},
and is a central concept of \cite[Chapter~8]{Apt-Prolog}.
(Programs/queries satisfying the condition are called there
{\em well-asserted}.)  
Formally, Th.\,\ref{th.correctness} is stronger than the corresponding
results in
 \cite{DBLP:conf/tapsoft/BossiC89}, or \cite{Apt-Prolog},
as they do not deal with calls and successes, 
or---respectively---with successes in the derivations.%
\footnote{%
  Thus the proof method of \cite[Chapter~8]{Apt-Prolog} is insufficient to
  obtain Lemma~\ref{lemma2}.
  However it can be used to obtain a weaker lemma,
  stating that the computed answers are in ${\it post}$.
  Such lemma is sufficient to derive Th.\,\ref{th.magic}.
}
 So we give its proof in the Appendix.

The method of Th.\,\ref{th.correctness} is a special case of that of
\cite{DM88}%
\footnote{%
In \cite{DBLP:journals/fac/AptM94} it is shown that the sufficient condition of
Th.\,\ref{th.correctness} is a special case of that of
\cite{DM88}.
}.
The main difference
is that call-success specifications in \cite{DM88} are not required to be
closed under substitution.
Another correctness proof methods 
for non-declarative properties,
with specifications not necessarily closed under substitution, are presented in
\cite{DBLP:conf/iclp/ColussiM91,moje.floyd-hoare}. 

Often we are interested in declarative properties of programs.
For such properties a simpler proof method exists,
usually attributed to \cite{Clark79}.
We illustrate that method in \ref{appendix2} by another proof of Corollary
\ref{corollary.lemma}. 
The reader is referred to
\cite[Sections 3.1, 3.2]{DBLP:journals/tplp/DrabentM05}
for a presentation, further references, and for a comparison with
methods dealing with non-declarative properties.

\appendix

\section{}
\label{appendix}
Here we present a formal definition of procedure calls and successes,
and a soundness proof for the method of proving programs correct w.r.t.\ 
call-success specifications (Th.\,\ref{th.correctness}).
The definition follows that of \cite{DM88}.

\begin{definition}[Calls and successes]
\label{def.call.success}
\rm
Let $Q_0,Q_1,Q_2,\ldots$ be the sequence of queries
and $\theta_1,\theta_2,\ldots$ the sequence of mgu's 
of an LD-derivation $\DD$.
Let $\theta_{i,j}= \theta_{i+1}\cdots\theta_j$ for $i< j$.

An atom $A$ is a {\em procedure call} in  $\DD$ iff
$A$ is the first atom of some $Q_i$
($Q_i = A,{\bf B}$).

\pagebreak[3]
An atom $A'$  is a {\em procedure success} (of a call $A$)
in  $\DD$
iff

-- $Q_i = A,{\bf B}$ for some $i\geq0$,

-- $Q_j = \mathbf{B}\theta_{i,j}$ for some $j> i$,

-- and $A' = A\theta_{i,j}$ for the least such $j$.
\end{definition}

\noindent
Notice that if $A'$ is a success of a procedure call $A$
(in an LD-derivation for a program $P$) then $A'$ is a computed answer
for $A$ (and $P$).
The corresponding successful derivation for $A$ can be constructed out of
the queries $Q_i,\ldots,Q_j$ as above,
by removing 
$\mathbf{B}\theta_{i,l}$ from each query
$Q_l = Q_l',\mathbf{B}\theta_{i,l}$,
for $l=i\ldots, j$
(where $\theta_{i,i}$ stands for $\epsilon$, and
$Q_i' = A$).

\medskip\noindent
PROOF of Theorem \ref{th.correctness}:
Assume that the conditions of the theorem are satisfied,
and consider an LD-derivation for $P$ and $Q$.
By \cite[Corollary 8.8]{Apt-Prolog},
each procedure call in the derivation is in ${\it pre}$.

As explained above, each procedure success $A'$ of a call $A$ is a
computed answer for $A$.  
By \cite[Corollary 8.9]{Apt-Prolog}
the computed answer is in ${\it post}$.
\hfill $\Box$

\noindent
\section{Declarative proof of Corollary \ref{corollary.lemma}}
\label{appendix2}

The proof method \cite{Clark79} 
is based on a property that, given an interpretation $I$,
if $I\models P$ then $I\models Q$ for each answer $Q$ of a program $P$.
Such $I$ is treated as a specification;
 $I\models P$ is a sufficient condition for correctness of $P$ w.r.t.\ $I$.

We will use term interpretations \cite[Section 4.4]{Apt-Prolog};
their interpretation domain is the set
of all the terms (of the given language).
Ground terms are interpreted as themselves.
A valuation for variables is a substitution.
Under a valuation $\eta$, a term $t$ is interpreted as $t\eta$.
An interpretation is (represented as) a set of atoms.
An atom $A$ is true in an interpretation $I$ under a valuation $\eta$ iff 
$A\eta\in I$.  Thus  $I\models A$ iff each instance of $A$ is in $I$.
For a clause $C = H\gets \seq B$ we have:
$I\models H\gets \seq B$ iff 
$B_1\eta,\cdots,B_n\eta\in I$ implies $H\eta\in I$
for each instance $C\eta$ of~$C$.

\newcommand{\MP}{{\it MP}}

\medskip\noindent
PROOF (of Corollary \ref{corollary.lemma}): \
Let us abbreviate $\MP = \magic(P,Q)$.
As a specification for $P$ we take the interpretation
\[
I = \{\, A \mid
A \mbox{ is an atom, }
\MP\notmodels \pre A \mbox{ or }
\MP\models A
\,\}.
 \]
Obviously:
\begin{equation}
\label{property.I}
\mbox{\color{black} If $A\in I$ then $\MP\models\pre A$ implies $\MP\models A$}.
\end{equation}
\pagebreak[3]

We show $I\models P$
(hence $P$ is correct w.r.t.\ $I$). 
Let $H\gets\seq B\in P$.  Assume $B_1\eta,\ldots,B_n\eta\in I$.
We have to show that $H\eta\in I$.
Notice first that
$
    \MP\models \pre H\eta, \
    \MP\models B_1\eta,\
    \ldots,
$
\linebreak[3]
$
    \MP\models B_{i-1}\eta
$\linebreak[3]
imply
$\MP\models\pre B_i\eta$
(by a clause of $\MP$ from case
\ref{magic.new.clauses} of Def.\,\ref{def.magic}),
and hence 
$\MP\models B_i\eta$, by (\ref{property.I}).
By simple induction we obtain that $\MP\models\pre H\eta$ implies
$ \MP\models B_1\eta,\ldots,\MP\models B_n\eta$,
and thus it implies
$\MP\models H\eta$ (by the clause from case
\ref{magic.modified.clauses} of Def.\,\ref{def.magic}).
If $\MP\notmodels\pre H\eta$ then
$H\eta\in I$ (by the definition of $I$).
Otherwise, by the implication above, $\MP\models H\eta$; thus $H\eta\in I$.
{\sloppy\par}

By the assumption of the Corollary,  $Q\sigma$ is an answer for $P$.
Thus from $I\models P$ it follows that $I\models Q\sigma$, hence $Q\sigma\in I$.
As $\MP\models\pre Q\sigma$, we have $\MP\models Q\sigma$.
\hfill $\Box$

\bibliographystyle{acmtrans}
\bibliography{bibmagic}

\begin{thebibliography}{}

\bibitem[\protect\citeauthoryear{Apt}{Apt}{1997}]{Apt-Prolog}
{\sc Apt, K.~R.} 1997.
\newblock {\em From Logic Programming to {P}rolog}.
\newblock International Series in Computer Science. Prentice-Hall.

\bibitem[\protect\citeauthoryear{Apt and Marchiori}{Apt and
  Marchiori}{1994}]{DBLP:journals/fac/AptM94}
{\sc Apt, K.~R.} {\sc and} {\sc Marchiori, E.} 1994.
\newblock Reasoning about {Prolog} programs: From modes through types to
  assertions.
\newblock {\em Formal Asp. Comput.\/}~{\em 6,\/}~6A, 743--765.

\bibitem[\protect\citeauthoryear{Beeri and Ramakrishnan}{Beeri and
  Ramakrishnan}{1991}]{DBLP:journals/jlp/BeeriR91}
{\sc Beeri, C.} {\sc and} {\sc Ramakrishnan, R.} 1991.
\newblock On the power of magic.
\newblock {\em J. Log. Program.\/}~{\em 10,\/}~1/2/3{\&}4, 255--299.

\bibitem[\protect\citeauthoryear{Bossi and Cocco}{Bossi and
  Cocco}{1989}]{DBLP:conf/tapsoft/BossiC89}
{\sc Bossi, A.} {\sc and} {\sc Cocco, N.} 1989.
\newblock Verifying correctness of logic programs.
\newblock In {\em TAPSOFT, Vol.2}, {J.~D\'{\i}az} {and} {F.~Orejas}, Eds.
  Lecture Notes in Computer Science, vol. 352. Springer, 96--110.

\bibitem[\protect\citeauthoryear{Bossi, Gabbrielli, Levi, and Martelli}{Bossi
  et~al\mbox{.}}{1994}]{s-semantics94}
{\sc Bossi, A.}, {\sc Gabbrielli, M.}, {\sc Levi, G.}, {\sc and} {\sc Martelli,
  M.} 1994.
\newblock The s-semantics approach: Theory and applications.
\newblock {\em J. Log. Program.\/}~{\em 19/20}, 149--197.

\bibitem[\protect\citeauthoryear{Clark}{Clark}{1979}]{Clark79}
{\sc Clark, K.~L.} 1979.
\newblock Predicate logic as computational formalism.
\newblock Tech. Rep. 79/59, Imperial College, London. December.

\bibitem[\protect\citeauthoryear{Colussi and Marchiori}{Colussi and
  Marchiori}{1991}]{DBLP:conf/iclp/ColussiM91}
{\sc Colussi, L.} {\sc and} {\sc Marchiori, E.} 1991.
\newblock Proving correctness of logic programs using axiomatic semantics.
\newblock In {\em Logic Programming, Proceedings of the Eigth International
  Conference}, {K.~Furukawa}, Ed. MIT Press, 629--642.

\bibitem[\protect\citeauthoryear{Deransart}{Deransart}{1993}]{DBLP:journals/tc%
s/Deransart93}
{\sc Deransart, P.} 1993.
\newblock Proof methods of declarative properties of definite programs.
\newblock {\em Theor. Comput. Sci.\/}~{\em 118,\/}~2, 99--166.

\bibitem[\protect\citeauthoryear{Drabent}{Drabent}{1997}]{moje.floyd-hoare}
{\sc Drabent, W.} 1997.
\newblock A {F}loyd-{H}oare method for {Prolog}.
\newblock {\em Link\"oping Electronic Articles in Computer and Information
  Science\/}~{\em 2}.
\newblock \url{http://www.ep.liu.se/ea/cis/1997/013/}.

\bibitem[\protect\citeauthoryear{Drabent and Ma{\l}uszy\'{n}ski}{Drabent and
  Ma{\l}uszy\'{n}ski}{1988}]{DM88}
{\sc Drabent, W.} {\sc and} {\sc Ma{\l}uszy\'{n}ski, J.} 1988.
\newblock {Inductive Assertion Method for Logic Programs}.
\newblock {\em Theoretical Computer Science\/}~{\em 59}, 133--155.

\bibitem[\protect\citeauthoryear{Drabent and Mi{\l}kowska}{Drabent and
  Mi{\l}kowska}{2005}]{DBLP:journals/tplp/DrabentM05}
{\sc Drabent, W.} {\sc and} {\sc Mi{\l}kowska, M.} 2005.
\newblock Proving correctness and completeness of normal programs -- a
  declarative approach.
\newblock {\em Theory and Practice of Logic Programming\/}~{\em 5,\/}~6,
  669--711.

\bibitem[\protect\citeauthoryear{Mascellani and Pedreschi}{Mascellani and
  Pedreschi}{2002}]{DBLP:conf/birthday/MascellaniP02}
{\sc Mascellani, P.} {\sc and} {\sc Pedreschi, D.} 2002.
\newblock The declarative side of magic.
\newblock In {\em Computational Logic: Logic Programming and Beyond}, {A.~C.
  Kakas} {and} {F.~Sadri}, Eds. Lecture Notes in Computer Science, vol. 2408.
  Springer, 83--108.

\bibitem[\protect\citeauthoryear{Nilsson}{Nilsson}{1995}]{DBLP:journals/tcs/Ni%
lsson95}
{\sc Nilsson, U.} 1995.
\newblock Abstract interpretation: A kind of magic.
\newblock {\em Theor. Comput. Sci.\/}~{\em 142,\/}~1, 125--139.

\bibitem[\protect\citeauthoryear{Nilsson and Maluszynski}{Nilsson and
  Maluszynski}{1995}]{nilsson.maluszynski.book}
{\sc Nilsson, U.} {\sc and} {\sc Maluszynski, J.} 1995.
\newblock {\em Logic, Programming and Prolog \/{\rm(2ed)}}.
\newblock Previously published by John Wiley \& Sons Ltd.
\newblock \url{http://www.ida.liu.se/~ulfni/lpp/}.

\bibitem[\protect\citeauthoryear{Ramakrishnan}{Ramakrishnan}{1991}]{DBLP:journ%
als/jlp/Ramakrishnan91}
{\sc Ramakrishnan, R.} 1991.
\newblock Magic templates: A spellbinding approach to logic programs.
\newblock {\em J. Log. Program.\/}~{\em 11,\/}~3{\&}4, 189--216.

\end{thebibliography}

\end{document}